\documentclass[11pt]{article}

\input epsf.sty
\usepackage{color} 
\usepackage[utf8]{inputenc} 
\usepackage{feynmp}

\usepackage{graphics,amsmath,amssymb}
\def\lromn#1{\uppercase\expandafter{\romannumeral#1}}

\begin{document}

\begin{flushright}
\end{flushright}

\begin{center}
\begin{Large}
\textbf{
Bifurcated symmetry breaking in scalar-tensor gravity
}
\end{Large}

\vspace{1cm}

M. Yoshimura

\vspace{0.5cm}
Research Institute for Interdisciplinary Science,
Okayama University \\
Tsushima-naka 3-1-1 Kita-ku Okayama
700-8530 Japan

\vspace{5cm}

{\bf ABSTRACT}

\end{center}

\vspace{1cm}

We present models that simultaneously predict  presence of
dark energy and cold dark matter along with slow-roll inflation.
The dark energy density is found to be of order
$( {\rm a \;few \;meV})^4$, and  the mass of  dark matter constituent is 
$\approx 1\,$ meV.
These numbers are given in terms of the present value of Hubble constant $H_0$
and the Plank energy $1/\sqrt{16 \pi G_N}$: they are $(H_0 M_{\rm P})^2$ 
for the energy density and
$(H_0 M_{\rm P})^{1/2}$ for the dark matter constituent mass.
The basic framework is a multi-scalar tensor gravity with non-trivial conformal coupling
to the Ricci scalar curvature in the lagrangian density.
The key  for a right amount of dark energy 
is to incorporate in a novel way
 the spatially homogeneous kinetic contribution of Nambu-Goldstone modes
in a spontaneously broken multi-scalar field sector.
Proposed theories are made consistent with general relativity tests
at small cosmological distances, yet are different from
general relativity at cosmological scales.
Dark matter is generated as spatially inhomogeneous component
of the scalar system, with roughly comparable amount to the dark energy.
In some  presented models a
cosmological bifurcation of symmetry breaking of scalar sector is triggered
by the spontaneous breaking of electroweak SU(2) $\times $ U(1)
gauge symmetry, hence the separation occurring simultaneously 
at the electroweak phase transition.
The best experimental method to test presented models
is to search for the fifth-force type of scalar exchange 
interaction with a force range, $O(10^{-2})$ cm, whose coupling to matter is
basically of gravitational strength.

\vspace{2cm}

Keywords
\hspace{0.5cm} 
Dark energy,
Dark matter,
Slow-roll inflation,
Scalar-tensor gravity,
Fifth force,
Electroweak phase transition

\newpage

\section
 {\bf Introduction}

General relativity (GR for brevity) is a remarkable success as  theory of gravity,
passing all stringent tests in the solar system, merging binary pulser and
gravitational wave detection \cite{gr tests}, \cite{gw and gr}, \cite{ns-merger gw}.
Nevertheless, inflation that solves
cosmological conundrums \cite{cosmology} requires a scalar degree of freedom 
\cite{inflation models 1}, \cite{inflation models 2},
and with introduction of conformal coupling the scalar degree of freedom
inevitably links to the gravitational tensor field.
Dark energy may also  be related to existence of
a scalar field \cite{dark energy scalar 1}, \cite{dark energy scalar 2}.

An interesting precursor of scalar-tensor gravity is Jordan-Brans-Dicke (JBD) theory 
\cite{jordan}, \cite{brans-dicke}
which however failed in many ways, and one of the reasons may be traced
to absence of scalar potential.
Scalar-tensor gravity with potential term incorporated  led to a plethora of
models which attempt to solve the dark energy problem.
All models of scalar-tensor gravity must clear classical and new tests of
general relativity, which turns out non-trivial in many proposed models.

In the present work we propose the idea of using, in a novel way, Nambu-Goldstone modes
of spontaneously broken symmetry in multi-scalar tensor gravity.
When kinetic contribution of these modes is  incorporated as
a part of effective scalar potential, it gives rise to a new type of repulsive forces
dependent on the metric determinant.
The scalar fields in cosmology may generate both spatially homogeneous and
and inhomogeneous components.
We show that the spatially homogeneous part behaves as dark energy, with
its equation of state factor -1, and the spatially inhomogeneous part behaves
as dark matter, with its number energy density 
decreasing  inversely proportional to the volume factor with cosmological evolution.
Their predicted energy densities at the present epoch
are of order $(H_0 M_{\rm P})^2 \approx ({\rm a \; few\; meV})^4$
(with $H_0 \approx 10^{-33}$ eV the Hubble energy and 
$M_{\rm P} =1/\sqrt{16\pi G_N}  \approx 10^{27}$ eV
the Planck energy in our definition).
Mass of dark matter constituent  is of order $(H_0 M_{\rm P})^{1/2} \approx\,$ a few meV,
hence all energy scales are given by this combination of Hubble and Planck energies.
Bifurcation of dark matter from dark energy may be triggered
by the electroweak SU(2) $\times $ U(1) gauge symmetry breaking.
Another great merit of Nambu-Goldstone modes
is that they clear GR tests in a trivial way.

The present paper is organized as follows.
In Section 2 we lay out our theoretical framework of scalar-tensor gravity
based on conformal coupling of multi-scalar sector to the massless graviton.
The scalar system is assumed to have a continuous global symmetry,
and role of Nambu-Goldstone modes is emphasized to produce 
the new type of repulsive forces.
In Section 3 we discuss cosmology of symmetric scalar-tensor gravity,
taking as an example the simplest O(2) model made of two real scalar fields,
and show that slow-roll inflation and late time accelerating universe
are both realized by a modest choice of parameter range.
Tracker solution at late times, recognized as spatially uniform,
 predicts a right amount of dark energy density
given by $O (H_0 M_{\rm P})^2 = O($1.7 meV$)^4$
 independent of model parameters.
Spatially inhomogeneous components are identified in
the linearized approximation around homogeneous solutions,
and are decomposed in terms of their
wave vectors, and each component is recognized as dark matter of definite momentum, with
its number density being inversely proportional to the volume factor.
Section 4 is devoted to explanation of the bifurcated symmetry breaking
triggered by the electroweak gauge symmetry breakdown.
Two possibilities exist depending on whether the scalar field has SU(2) $\times $ U(1)
quantum numbers or not.
It is found that dark scalars in viable models should  have no SU(2) $\times $ U(1) quantum number.
In Section 5 we present some rudimentary ideas on how this class of models may
be tested experimentally in laboratories on earth.
The paper ends with a brief summary.

We use the natural unit of $\hbar = c = 1$ and the unit Boltzmann constant
$k_B=1$  throughout the present work  unless otherwise stated.

\section
{\bf Theoretical framework}

We consider a class of local field theory in four space-time dimensions
of the following type:
\begin{equation}
{\cal L} = \sqrt{-g}\,\left( -{\rm M}_P^2 f(\varphi_i) R + \frac{1}{2}(\partial \varphi_i)^2
- V(\varphi_i) + {\cal L}_{m}(\psi) \right) \,,
\label {jbd frame}
\end{equation}
written in terms of graviton metric field $g_{\mu \nu}$, a number of scalar fields $\varphi_i$, and
standard model fields generically denoted by $\psi$,
and $(\partial \varphi_i)^2 = g_{\mu \nu} \partial^{\mu} \varphi_i \partial^{\nu} \varphi_i$.
Our signature convention of flat metric is $(1\,, -1\,, -1\,, -1)$), and the metric determinant $g < 0$.
The constant ${\rm M}_P^2 = 1/ 16\pi G_N \approx (1.72 \times 10^{18} {\rm GeV })^2 $
is our squared Planck energy.
The scalar potential $V(\varphi_i) $ and the conformal coupling function
$ f(\varphi_i)$ are assumed to have a global
symmetry, the simplest case being a function of field modulus, $\varphi^2=\sum_{i=1}^{N} \varphi_i^2 $,
for real $N$ scalars which gives $O(N)$ symmetry.
With a constant $f(\varphi_i)$ the theory is a trivial extension of
standard particle theory, and can realize inflation taking an appropriate form
of potential $V(\varphi_i)$.
We introduce here a non-trivial function $ f(\varphi_i)$, 
and  this function may be expanded in terms of 
dimensionless field $(\varphi_i/{\rm M}_P)^2$.

The metric frame thus introduced may be called JBD frame.
It is often more convenient to introduce the Einstein frame by
a Weyl rescaling, $\bar{g}_{\mu\nu} = f(\varphi) g_{\mu\nu}  $,
\begin{eqnarray}
&&
{\cal L}(\bar{g}_{\mu\nu},\varphi_i) = \sqrt{-\bar{g}} \left(-{\rm M}_P^2 R(\bar{g}_{\mu\nu})
+ \frac{5}{2 f} (\partial \varphi_i)^2 - \frac{1}{f^2} V(\varphi) 
 + \frac{1}{f^2}  {\cal L}_m (\psi, \frac{\bar{g}_{\mu\nu}}{f} )
\right)
\,.
\end{eqnarray}
We denote hereafter the new metric $\bar{g}_{\mu\nu}  $ by $g_{\mu\nu}  $.
A simple conformal factor and potential of the form is considered in the present work;
\begin{eqnarray}
&&
f(\varphi) = 1+ f_2 \frac{ \varphi^2}{M_{\rm P}^2} \,, \hspace{0.5cm} f_2 > 0
\,,
\label {conformal factor}
\\ &&
V(\varphi) = V_0 - \frac{m_{\varphi}^2}{2}  \varphi^2
+ \frac{\lambda}{4} ( \varphi^2)^2
\,, \hspace{0.5cm}
m_{\varphi}^2 > 0 
\,, \hspace{0.5cm}
\lambda > 0
\,.
\label {potential}
\end{eqnarray}
The potential of scalar system $ V(\varphi)$ exhibits a spontaneous symmetry breaking
due to a negative curvature at the origin $\varphi_i = 0$.
Our model is characterized by two dimensionless parameters, $\lambda\,, f_2$,
taken to be of order unity, and other parameters of non-vanishing mass
dimensions. The constant $V_0$ is later adjusted by requirement of
vanishing cosmological constant.
The mass value $m_{\varphi}$ (taken positive) is left arbitrary for the moment.
More complicated generalization of potential and conformal factor is possible, but this simplest case is sufficient
to explain main features of our theory.

Scalar kinetic term can be recast to the standard form by a change of
field normalization. First, we separate modulus part and angular part by
introducing orthogonal transformation $O,\, O^T O = O O^T = 1$,
which transforms  a reference vector,
say $(\varphi, 0, \cdots, 0)$, to general $N$-vector $\vec{\varphi} = (\varphi_i) $.
Then, the kinetic term is written, using $\varphi = \sqrt{\sum_i \varphi_i^2}$,
\begin{eqnarray}
&&
\sum_i (\partial \varphi_i)^2 = (\partial \varphi)^2 + \frac{1}{N} \varphi^2  
\sum_{ij} (\partial O_{ij})^2
\,.
\end{eqnarray}
The second term written in terms of derivatives of
orthogonal matrix elements is contribution from Nambu-Goldstone modes,
since the potential $V(\varphi) $ does not depend on Nambu-Goldstone angular mode variables that appear
in $O$.
In the simplest O(2) case this term is $\varphi^2 (\partial \theta)^2$
when $\vec{\varphi} = \varphi (\cos \theta, \sin \theta)$.
In O(3) case kinetic  Goldstone term is
$ \varphi^2 \left( \partial \theta^2 + \sin^2 \theta \partial \psi^2 \right)$
using the standard angular variables, $\vec{\varphi} = \varphi
(\sin \theta \cos \psi, \sin \theta \sin \psi, \cos \theta)$.
In both cases they are written in terms of generalized angular momentum operator
$\vec{L}^2$ with $\vec{L} = \vec{\varphi} \times \partial \vec{\varphi}$.
In cosmology $\partial $ is practically time derivative.

Except Nambu-Goldstone kinetic terms, one can change scalar field variables
to the standard form;
\begin{eqnarray}
&&
\phi =\sqrt{5} \int^{\varphi} \frac{d\varphi} { \sqrt{f(\varphi)}}
\,, \hspace{0.5cm}
\frac{5}{2 f} (\partial \varphi)^2 = \frac{1}{2} (\partial \phi)^2 
\,,
\\ &&
{\cal L}(g_{\mu\nu},\phi) = \sqrt{-g } \left(- {\rm M}_P^2 R(g_{\mu\nu})
+ \frac{1}{2 } (\partial \phi)^2  + GK - \frac{1}{f^2} V(\phi) 
+ \frac{1}{f^2}  {\cal L}_m (\psi, \frac{g_{\mu\nu}}{f} )
\right)
\,,
\\ &&
GK = \frac{1}{2N} \varphi^2  \sum_{ij} (\partial O_{ij})^2
\,.
\end{eqnarray}
For the choice of eqs.(\ref{conformal factor}), (\ref{potential}),
these are explicitly
\begin{eqnarray}
&&
\varphi = \frac{ M_{\rm P}}{\sqrt{f_2} }\sinh (\sqrt{\frac{ f_2}{5 }} \frac{\phi }{ M_{\rm P}} )
\,, \hspace{0.5cm}
f = \cosh^2 (\sqrt{\frac{ f_2}{5 }} \frac{\phi }{ M_{\rm P}} )
\,,
\\ &&
\frac{V}{f^2} = \frac{V_0}{f^2} + \frac{ \lambda}{ 4 f_2^2}  M_{\rm P}^4 ( 1 - \frac{1}{f} )^2
+ \frac{1}{2 f_2 }  M_{\rm P}^2 m_{\varphi}^2
( 1 - \frac{1}{f} + \frac{1}{f^2})
\,.
\label {potential/f2}
\end{eqnarray}
This has a limiting behavior at $\phi \rightarrow 0$,
(assuming $m_{\varphi} \ll M_{\rm P}$ and neglecting $O(\phi^2/M_{\rm P}^2 )$ terms)
\begin{eqnarray}
&&
\frac{V}{f^2} \rightarrow (\phi \;{\rm independent\; terms\,} ) - 
\frac{1}{10}  m_{\varphi}^2  \phi^2 + \frac{\lambda}{100} \phi^4
\,.
\end{eqnarray}

Nambu-Goldstone modes appear only in kinetic terms of lagrangian.
The field equation for O(2) Nambu-Goldstone mode is 
$\partial (\sqrt{-g}\, \phi^2 \partial \theta ) = 0 $,
to give its generic solution,
\begin{eqnarray}
&&
\sqrt{-g} \phi^2 (\partial \theta)^2 =  \frac{c^2}{ \sqrt{-g} \, \phi^2} 
\,,
\end{eqnarray}
with $c = \sqrt{-g} \phi^2 \partial \theta $ a constant of integration
giving the value of angular momentum operator.
One may define an effective potential, adding a centrifugal term;
in the small field limit,
\begin{eqnarray}
&&
\hspace*{-1cm}
V_{{\rm eff}}(\phi) = \frac{V(\phi)  }{f^2} + \frac{c^2}{ -g\, \phi^2}  \approx V_0 
- \frac{1}{10} m_{\varphi}^2 \phi^2 + \frac{\lambda}{100} \phi^4
+\frac{c^2}{ -g\, \phi^2} 
\,, \hspace{0.5cm}
\phi^2 = \sum_{i=1}^N \phi_i^2
\,.
\end{eqnarray}
The centrifugal term changes location of potential minimum
(at $\sqrt{10/\lambda} \,m_{\varphi} $ without centrifugal term).
Except in inflationary epoch this potential formula and its extensions in the low energy limit
are adequate in our discussion.

Nambu-Goldstone kinetic terms in extended models may emerge in a far more non-trivial way.
Suppose, taking a real $N$ component model, that two parts of $n$ and $m= N - n$
components separately appear with two different integration constants, 
\begin{eqnarray}
&&
\frac{c_t^2}{ -g\, \phi_t^2}
\,, \; 
( \phi_t^2 = \sum_{i=1}^n \phi_i^2)
\,, \hspace{0.5cm}
{\rm and} \hspace{0.5cm}
 \frac{c_l^2}{ -g\, \phi_l^2}
\,, \; 
( \phi_l^2 = \sum_{i=n+1}^N \phi_i^2 )
\,, \hspace{0.5cm}
c_t \neq c_l 
\,,
\end{eqnarray}
giving different centrifugal repulsions.
This occurs when the scalar system has spontaneously a broken O($n$) $\times $ O($m$)
symmetry.
We later encounter an example of  $m=1, c_l = 0$.
This is a bifurcation of symmetry breaking in the dark sector, and provides
a separation mechanism of dark energy in the dark sector.

We note in passing that there exist stringent constraints from
GR tests imposed  on an interesting model of restored discrete symmetry
\cite{symmetron model}, although the necessary Vainshtein decoupling \cite{vainshtein}
around a massive astronomical body is realized  in this and related models.
Historically, the Vainshtein decoupling played an important role in the massless limit
of massive spin 2 graviton theory.
Two degrees of freedom associated with the massless graviton decouples
from the rest of three degrees of freedom within what is called Vainshtein radius around
a massive astronomical object,
becoming of order $1.9 \times 10^{28}\,  $cm (roughly the size of
presently observable universe) for the sun and a graviton mass of order the Hubble constant $H_0
\approx 10^{-33}$ eV.
(This radius changes as $\propto m_g^{ -4/5}$ with the graviton mass $ m_g$.)

A  different decoupling mechanism works in our
scalar-tensor gravity due to centrifugal repulsive potential of Nambu-Goldstone modes, 
which is a key in passing GR tests at smaller distances.

\section
{\bf Cosmology}

We first discuss cosmology in O(2) symmetric scalar-tensor theory.
O($N \geq 3$) symmetry extension is straightforward.
The spatially flat Robertson-Walker metric \cite{cosmology} given by 
$ds^2 = dt^2 - a^2(t) \vec{dr}^2 $ is sufficient for our discussion.
In the flat spacetime the absolute value of scale factor $a(t)$
may be taken arbitrarily,
and we  choose the normalized scale factor such that
the present scale factor is unity: $a(t_0) = 1$.
The  Nambu-Goldstone $\theta$ field equation is
\begin{eqnarray}
&&
\frac{d}{dt} ( a^3 \phi^2 \frac{d \theta}{dt}) = 0
\,,
\end{eqnarray}
which makes it possible to incorporate in the lagrangian density for scalar field $\phi$
a centrifugal term, to give an effective potential $V_{{\rm eff}}(\phi) $,
\begin{eqnarray}
&&
{\cal L}_{\phi} = a^3 \left( \frac{1 }{2}\dot{\phi}^2 
- V_{{\rm eff}}(\phi) 
\right)
\,, \hspace{0.5cm}
V_{{\rm eff}}(\phi) = \frac{V(\phi)}{f^2}  + \frac{c^2}{2 a^6 \, \phi^2} 
\,,
\end{eqnarray}
with $c > 0$ an integration constant. 
$c$ is related to the Hubble constant $H_0$ when we discuss late time solution.
The Einstein and scalar field equations are then 
\begin{eqnarray}
&&
(\frac{\dot{a}}{a})^2 = \frac{8\pi G_N}{3} (\rho_{\phi} + \rho_m)
\,, \hspace{0.5cm}
\rho_{\phi} =  \frac{1}{2}( \dot{\phi}^2 + V_{{\rm eff}} )
\,, \hspace{0.5cm}
\rho_m = T^m_{00}
\,,
\\ &&
\ddot{\phi} + 3 \frac{\dot{a}}{a} \dot{\phi} +
\partial_{\phi} V_{{\rm eff}}(\phi) =  
- \, \frac{\partial_{\phi} f}{2 f}  T^m
\,,
T^m = g^{\mu \nu} T^m_{\mu \nu}
\,,
\\ &&
T^m_{\mu\nu}
=  \frac{2}{\sqrt{-g} f^2}  \frac{\delta }{\delta g^{\mu\nu}}
 \left(\sqrt{-g}\, {\cal L}_m (\psi, \frac{g_{\mu\nu}}{f} ) \right)
\,,
\end{eqnarray}
dot being time derivative.
We assume for this discussion as if a non-vanishing $c$ existed since a pre-inflation epoch.
But at inflationary epoch and also in subsequent radiation-dominated epoch the centrifugal term
is not important, and its importance becomes evident at late epoch of accelerating universe.
For radiation and matter energy density $\rho_m$ we only need to know that
$\rho_m \propto 1/a^4$ in radiation dominated epoch
(also $T^m = 0$), while $\rho_m \propto 1/a^3$
in matter dominated epoch.

\begin{figure*}[htbp]
 \begin{center}
 \epsfxsize=0.6\textwidth
 \centerline{\epsfbox{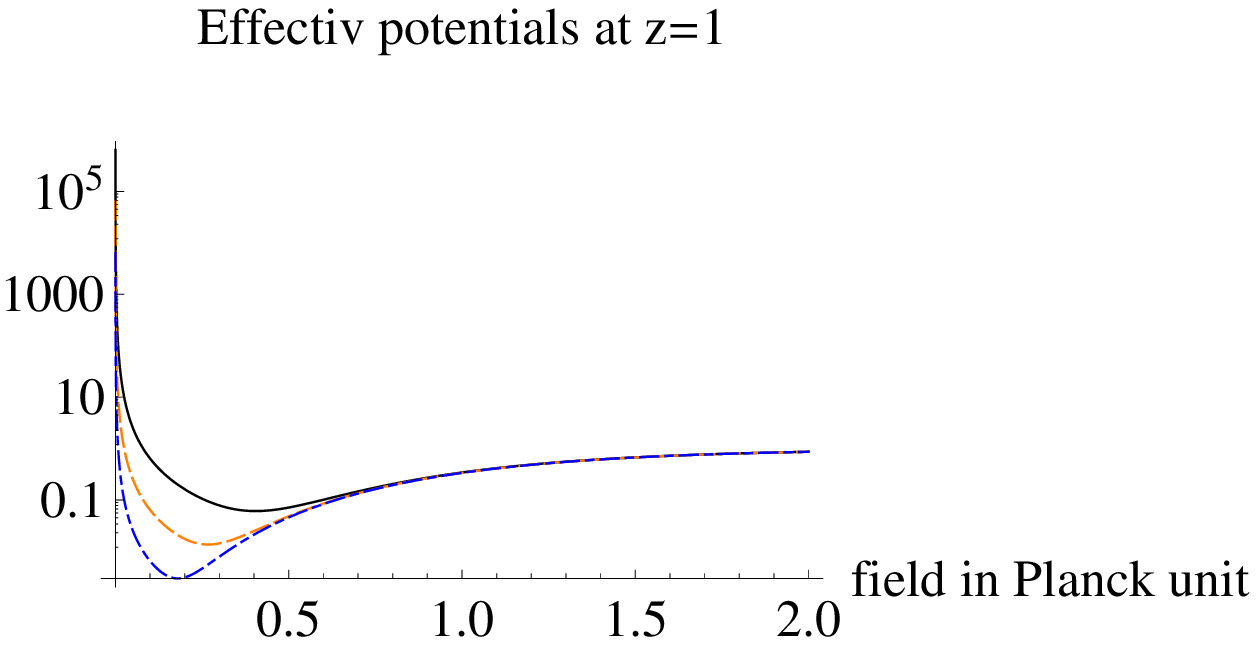}} \hspace*{\fill}\vspace*{1cm}
   \caption{
Effective potential $v_{{\rm eff}} (X,a) $ of eq.(\ref {dimless potential}) at the present redshift $z=1\,, a=0.5$ 
for three coefficients, $(k_1, k_2) = 1\,, 10^{-6}$ and $k_3= (10^{-4}, 10^{-5}, 10^{-6})$,
in solid black, dashed red, and dash-dotted blue, respectively.
}
   \label {effective potential}
 \end{center} 
\end{figure*}

We fine tune the cosmological constant such that
$V_0 + m_{\varphi}^2 M_{\rm P}^2/ (2f_2 )=0 $ (the second term
arising from the zero field limit of $\propto 1-1/f+1/f^2 $ term in eq.(\ref{potential/f2})).
It is suitable to use dimensionless  Planckian time unit 
$\tau = M_{{\rm P}} t$ and rewrite gravitational and field equations:
\begin{eqnarray}
&&
\frac{1}{a^2} ( \frac{d a}{d \tau})^2 = \frac{ 1}{6}
\left( \frac{5}{ 2 f_2} (\frac{dX}{d\tau})^2 + v_{{\rm eff}} (X,a) 
\right)
\,, \hspace{0.5cm}
X = \sqrt{\frac{f_2 }{5 }} \frac{\phi }{M_{\rm P} }
\,,
\label {dimless einstein eq}
\\ &&
\frac{d^2 X}{d\tau^2} + 3 \frac{1}{a} \frac{d a}{d \tau} \frac{dX}{d\tau} +
\partial_X v_{{\rm eff}} (X,a) = 0
\,,
\label {dimless scalar eq}
\\ &&
v_{{\rm eff}} (X,a ) \equiv \frac{V_{{\rm eff}}  }{ M_{\rm P}^4}
 = k_1 ( 1 - \frac{1}{f(X)})^2 + k_2 ( 1 - \frac{1}{f(X)} + \frac{1}{f^2(X)}) + \frac{k_3 }
{a^6\, X^2}
\,, \hspace{0.5cm}
f(X) = \cosh^2 X
\,, 
\label {dimless potential}
\\ &&
k_1 = \frac{\lambda}{4f_2^2}
\,, \hspace{0.5cm}
k_2 = \frac{1 }{2f_2 } \frac{m_{\varphi}^2 }{M_{\rm P}^2}
\,, \hspace{0.5cm}
k_3 =  \frac{f_2\, c^2 }{10   M_{\rm P}^6 } 
\,.
\label {potential parameters}
\end{eqnarray}
The dimensionless effective potential, $v_{{\rm eff}} (X,a) $ of eq.(\ref{dimless scalar eq}), 
near the present cosmological epoch at a redshift 1 or $a=1/2$ (in our normalized unit)
is illustrated in Fig(\ref{effective potential}),
although the chosen parameters (for a less computer tension) are not appropriate for a realistic choice.
Nonetheless, solutions derived from differential equations for graviton and
scalar field exhibit a rich variety of interesting behaviors.
Thus, this model is capable of solving cosmological conundrums both
at inflation and at late times.

\subsection
{\bf Inflationary epoch ending at thermalized hot universe}

Inflation in the usual picture occurs  when the potential dominates over kinetic term,
$ V_{{\rm eff}} \gg \dot{\phi}^2 $.
Suppose that a large field $\phi \geq M_{{\rm P}}$, or $X \geq 1$, exists in a
pre-inflationary epoch, presumably due to fluctuation of quantum gravity effects.
The slow-roll inflation \cite{inflation models 1}, \cite{inflation models 2}, \cite{cosmology}
is realized under the two conditions, $|V_{{\rm eff}}'/V_{{\rm eff}} | \ll  1/M_{\rm P} $
and $| V_{{\rm eff}}''/V_{{\rm eff}} | \ll 3/(2 M_{\rm P}^2) $
(the prime $'$ indicating field derivative), 
which read in our model as
\begin{eqnarray}
&&
| \frac{\partial_X v_{\rm eff} }{ v_{\rm eff}  } | \ll \sqrt{\frac{ 5}{ f_2}} 
\,, \hspace{0.5cm}
| \frac{ \partial_X^2 v_{\rm eff} }{ v_{\rm eff} } | \ll \frac{15 }{ 2f_2}
\,.
\label {slow-roll conditions}
\end{eqnarray}
We expect that inflation ends at $X= X_f \leq O(1)$.

With $f_2 = O(1)$, the first condition of eq.(\ref{slow-roll conditions}) turns out more stringent
than the second,
and one derives a sufficient condition, 
which becomes a condition for initial values of $aX $, assuming
tracker solution later derived (see  $c$ of eq.(\ref{c value})\,),
\begin{eqnarray}
&&
\frac{2k_3 }{a^3 (a X)^3 } \ll \sqrt{\frac{5 }{ f_2}} \left( k_1 + 
\frac{k_3 }{a^3 (a X)^3 } \right)
\,.
\end{eqnarray}
This is a condition readily satisfied, for instance by taking
$f_2 \ll \frac{5}{4}$.

\begin{figure*}[htbp]
 \begin{center}
 \epsfxsize=0.6\textwidth
 \centerline{\epsfbox{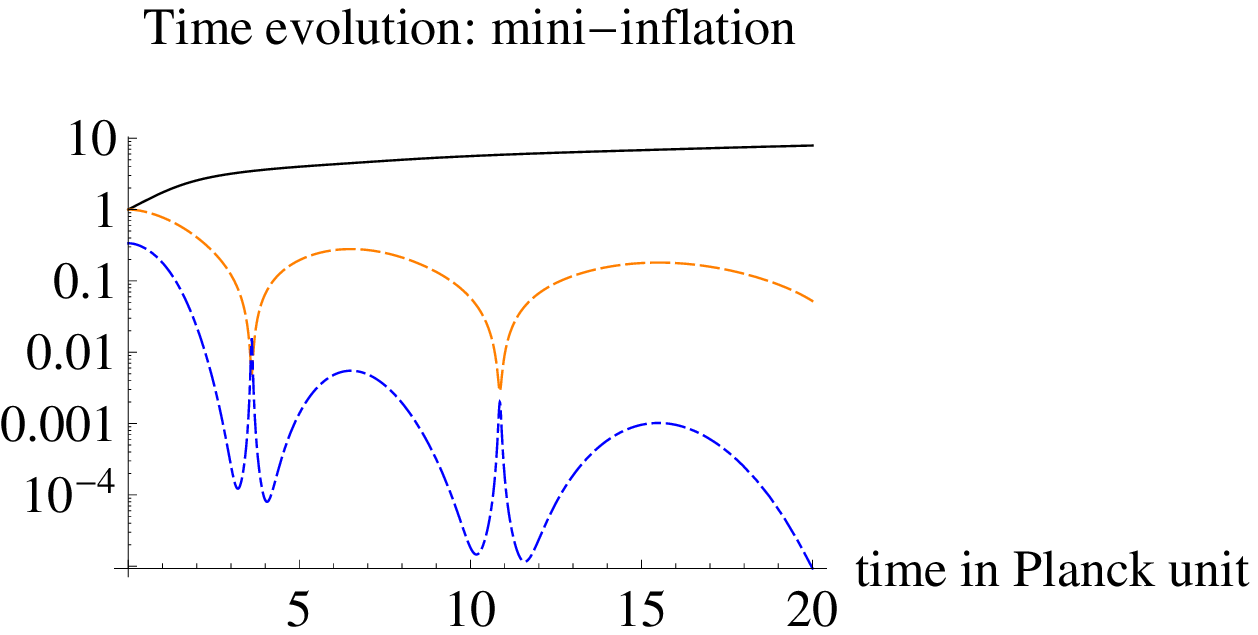}} \hspace*{\fill}\vspace*{1cm}
   \caption{
Time evolution solving dimensionless equations,
eq.(\ref{dimless einstein eq}) and (\ref{dimless scalar eq}) in logarithmic scale:
Scale factor  $a$ in solid black, scalar field $X $ in in dashed red,
potential $ v_{{\rm eff}} (X,a ) $ in dash-dotted blue.
Parameters used in the dimensionless potential, eq.(\ref{dimless potential}), are
$k_1 = 1\,, k_2= 10^{-6}\,, k_3 = 10^{-5} $ and initial condition 
$a(0)=0.5\,, \phi(0)=1\,, \dot{\phi}(0) = 0$ is assumed.
}
   \label {time evolution during inflation 1}
 \end{center} 
\end{figure*}

We illustrate in  Fig(\ref{time evolution during inflation 1}) a toy model example of time evolution.
Initial condition for numerical simulations has been determined by
seeking initial values $a(0)\,, \phi(0)$ when one imposes $v_{{\rm eff}} = O(1)$
and $\partial_{\phi}v_{{\rm eff}} \approx 0$.

Thermalized universe is expected to emerge using  production mechanism of
standard model particles as outlined in \cite{preheating 1}, \cite{preheating 2}.
Rapidly oscillating scalar field $\phi(t)$ at the end of inflation gives rise to copious Higgs boson production through coupling to standard model Higgs  boson of the form,
$\xi_{\phi H}\, \vec{\phi}\,^2 H^{\dagger} H $.
Non-perturbative parametric amplification of Floquet-type is responsible for this mechanism.
Thermal energy of produced Higgs pairs is then converted to other
relativistic particles, which achieves radiation dominated universe.

\subsection
{\bf Epoch of accelerating universe again}

Energy density of thermalized relativistic and non-relativistic particles changes
with cosmic scale factor $a(t)$ as $\propto 1/a(t)^4$ and $\propto 1/a(t)^3$,
hence radiation-dominated universe appears at earlier epoch  and then matter dominated-universe follows.
Dark energy density does not change with expansion, and
the universe at later epoch may enter into dark energy dominated universe governed by the
scalar field.
Note that the settled potential minimum at the end of inflation
differs from the true minimum at later epochs, and this mismatch accelerates
the late time universe.

Detailed scalar dynamics at redshift $< 10^3$ is analyzed by solving the equation,
\begin{eqnarray}
&&
\ddot{\phi} + 3 \frac{\dot{a}}{a} \dot{\phi} - \frac{m_{\varphi}^2}{5} \phi
+ \frac{\lambda}{ 25 } \phi^3 - \frac{c^2}{a^6 \phi^3} 
 = - \sqrt{\frac{ f_2}{5 }} \frac{\rho_m}{M_{\rm P} }
\,,
\\ &&
(\frac{\dot{a}}{a})^2 = \frac{1}{ 6 M_{\rm P}^2} ( \frac{\dot{\phi}^2}{2} 
+ V_{{\rm eff}} + \rho_m )
\,, \hspace{0.5cm}
V_{{\rm eff}} = \frac{\lambda}{100} \phi^4 - \frac{m_{\varphi}^2}{10} \phi^2
+ \frac{c^2}{2 a^6 \phi^2} 
\,,
\end{eqnarray}
where  the dark matter energy density $\rho_m$ is estimated as
$(2.2\, {\rm meV})^4 h^2/a^3\,, h \sim 0.7 $, from the present observed value.

We shall search solutions in which matter contribution $\rho_m$ is negligible.
Assuming a small $m_{\varphi}^2 $, we seek a solution in which
two terms, quartic field term and centrifugal repulsion, nearly balance.
This gives 
\begin{eqnarray}
&&
(a \phi)^6 = \frac{25}{\lambda} c^2
\,.
\end{eqnarray}
Inserting this relation into the Einstein equation, one derives
\begin{eqnarray}
&&
(a\,\dot{a})^2 = \frac{ c^{ 2/3}}{ 8 M_{\rm P}^2} (\frac{\lambda}{25})^{1/3}
\,.
\end{eqnarray}
Under the condition $a(t_0) = 1$, the solution is given by
\begin{eqnarray}
&&
a(t) = \sqrt{\frac{t+ t_i }{ t_0+ t_i}}
\,, \hspace{0.5cm}
\phi(t) = ( \frac{25}{\lambda} c^2 )^{1/6}\sqrt{\frac{ t_0+ t_i}{t+ t_i }}
\,.
\end{eqnarray}
The present Hubble constant $H_0$ is defined by $(da/adt)_{t = t_0} = H_0 $,
which determines $t_i$; $t_0 + t_i = (2H_0)^{-1} $.
From observational values one may estimate 
$t_i \approx 0.67 \times 10^{10}$ years
(taking $h=0.7$) compared to $t_0 \approx 1.36 \times 10^{10}$ years.
In order to determine $c$, we go back to the Einstein equation, which reads as
\begin{eqnarray}
&&
\frac{1 }{4(t+ t_i)^2 } = \frac{c^{ 4/3}}{ 8 M_{\rm P}^2} (\frac{\lambda}{25})^{1/3} 
\frac{1 }{ 4H_0^2 (t+ t_i)^2 } 
\,.
\end{eqnarray}
Thus,
\begin{eqnarray}
&&
c^{1/3} =2^{3/4} (\frac{\lambda}{25})^{-1/12} (H_0 M_{\rm P})^{1/2}
\simeq 3.7 \times 10^{-3}\, {\rm eV} \lambda^{-1/12}
\,.
\end{eqnarray}
The equation of state factor $w_{\phi}$ is defined by
\begin{eqnarray}
&&
w_{\phi} \equiv \frac{\frac{\dot{\phi}^2 }{ 2} - V_{{\rm eff}}}
{ \frac{\dot{\phi}^2 }{ 2} + V_{{\rm eff}}}
\,,
\end{eqnarray}
and is found to be nearly $-1$, because the ratio
\begin{eqnarray}
&& 
\frac{ \dot{\phi}^2}{ 2 V_{{\rm eff}} } = 
\frac{1}{6\sqrt{2}} (\frac{\lambda}{25})^{-1/2} \frac{H_0}{M_{\rm P}} \frac{1}{H_0( t+ t_i )}
\,,
\end{eqnarray}
is small with $H_0/M_{\rm P} \approx 10^{-60} $ for the solution we found.
We call this solution the tracker solution.

The dark energy density is independent of unknown couplings 
$\lambda\,, f_2\,, m_{\varphi} $ of the model, and is given by
\begin{eqnarray}
&&
V_{{\rm eff}}(t)  =  \frac{3}{4} (H_0 M_{\rm P})^2 (\frac{t_0 + t_i}{ t + t_i})^2
\sim  ( 1.6 \, {\rm meV})^4 (\frac{t_0+ t_i }{ t+ t_i})^2
\,.
\end{eqnarray}
This is a prediction of tracker solution for a single Goldstone mode.
Its value can be raised to $N$ times this value when there exist $N$ Goldstone modes.

To check how this solution is good, we analyze the linearized equation around this solution.
In terms of $\delta \phi = \phi - \phi_{T}$, the linearized equation is
\begin{eqnarray}
&&
\frac{d^2 \delta \phi}{dt^2} + \frac{3}{2 t} \frac{d \delta \phi}{dt} + \frac{K}{t} \delta \phi = 0
\,, \hspace{0.5cm}
K = 6 \sqrt{2} (\frac{\lambda}{25})^{1/2} M_{\rm P}
\,.
\end{eqnarray}
We have numerically integrated this equation in a time range 
$K t = 10^3 \sim 10^5$, and found that $|\delta \phi| \propto 1/\sqrt{t}$.
Hence deviation from the tracker solution is a minor correction.

We now analyze spatially inhomogeneous contribution of scalar field
by working out linearized field equation around the tracker solution.
Differential equations to be solved are
\begin{eqnarray}
&&
\left( \frac{\partial^2}{\partial t^2} + 3 \frac{\dot{a}}{a} \frac{\partial}{\partial t}
- \frac{1}{a^2} \vec{\nabla}^2 
\right) \delta \phi_t = 
- \left(\frac{\partial^2 V_{\rm eff} }{ \partial \phi_t^2}\right)_{\rm Tracker}
\delta \phi_t   
\,.
\end{eqnarray}
Coefficient of right hand side is calculated as
\begin{eqnarray}
&&
\frac{\partial^2 V_{\rm eff} }{ \partial \phi_t^2} =
2^{3/2} (\frac{ \lambda }{ 25})^{1/2 } H_0 M_{\rm P}
\simeq (1.26\, {\rm meV})^2 \lambda^{1/2 }
\,,
\end{eqnarray}
at the present epoch $t=t_0$.

The emerging picture of this simplest two-component O(2) model is that
Nambu-Goldstone centrifugal repulsion gives rise to a homogeneous
dark energy with its energy density of order $(H_0 M_{\rm P})^2$, 
while inhomogeneous part provides dark matter of mass
 $\approx \lambda^{1/4} (H_0 M_{\rm P})^{1/2}$.

\section
{\bf Extension for bifurcated symmetry breaking}

\subsection
{\bf Trigger by  spontaneous electroweak symmetry breaking}

There is no reason the scalar sector is not linked to the Higgs doublet present in
SU(2) $\times $ U(1) electroweak theory spontaneously broken down to U$_{\rm EM}$(1).
There are two possibilities, depending on whether dark scalars have SU(2) $\times $ U(1)
quantum numbers or not.

We first discuss the case of dark scalar being a SU(2) $\times$ U(1) singlet.
In this subsection we neglect temperature effects, and
concentrate on tracker solutions at temperatures much below the electroweak phase 
transition. In the next subsection we consider the region around the electroweak
critical temperature.
The potential part of fundamental lagrangian is given by
\begin{eqnarray}
&&
V(H, \phi) = \lambda_{\phi} (\phi^2 - \frac{m_{\phi}^2}{4 \lambda_{\phi}})^2
+ \lambda_H (H^{\dagger} H - \mu_H^2 )^2 + \xi_{\phi H}\,\phi^2 H^{\dagger} H
\,, \hspace{0.5cm}
\phi^2 = \sum_{i=1}^N \phi_i^2
\,,
\end{eqnarray}
with $H = (H_1 + i H_2, H_3 + i H_4)$ using real fields $H_i,\, i = 1 \cdots 4$.
By a SU(2) $\times$ U(1) transformation one can take homogeneous fields
parallel to $H = (H_1, 0)$.
The effective Higgs-scalar potential  is given by
\begin{eqnarray}
&&
\hspace*{-1cm}
V_{\rm eff}(H_1, \phi_1, \phi_t) =  
\lambda_{\phi} (\phi_1^2 + \phi_t^2 - \frac{m_{\phi}^2}{4 \lambda_{\phi}})^2
+ \lambda_H (H_1^2 - \mu_H^2 )^2 + \xi_{\phi H}\,(\phi_1^2 + \phi_t^2)  H_1^2  
+ \frac{c^2}{ 2 a^6 \phi_t^2}
\,, \hspace{0.3cm}
\phi_t^2 = \sum_{i=1}^{N-1} \phi_i^2
\,.
\label {new higgs potential 3}
\end{eqnarray}
It would be conceptually simpler to think of a O(N-1) broken scheme in which the
potential has a form of linear combinations among
 $ (\phi_t^2 - \mu_t^2)^2, \, (\phi_1^2 - \mu_1^2)^2, \, \phi_t^2 \phi_1^2,\, \phi_t^2,\,
\phi_1^2$ with different coefficients.
The potential given by eq.(\ref{new higgs potential 3}) is a special  combination
of minimal different coefficients.
It turns out that this simplified form is sufficient for our purpose.

Potential minimum is determined by solving algebraic equations for $H_1\,, \phi_1\,, \phi_t $,
\begin{eqnarray}
&&
2 \lambda_H (H_1^2 - \mu_H^2) + \xi_{\phi H} (\phi_1^2 + \phi_t^2)= 0
\,, \hspace{0.5cm}
\phi_1 ( \phi_1^2 + \phi_t^2 + \xi_{\phi H} H_1^2 - \frac{m_{\phi}^2}{4 \lambda_{\phi}}) = 0
\,,
\\ &&
2 \phi_t^4 \left( 2\lambda_{\phi} (\phi_1^2 + \phi_t^2 - \frac{m_{\phi}^2}{4 \lambda_{\phi}} ) 
+ \xi_{\phi H} H_1^2 \right) = \frac{c^2 }{a^6 }
\,.
\end{eqnarray}
Assuming a small $m_{\phi}^2 $, 
an interesting tracker solution satisfies
\begin{eqnarray}
&&
H_1^2 = \mu_H^2 - \frac{ \xi_{\phi H} }{2\lambda_{H} } \phi_t^2
\,, \hspace{0.5cm}
2 \phi_t^4 \left( ( 2\lambda_{\phi} - \frac{ \xi_{\phi H}^2 }{2\lambda_{H} }) \phi_t^2  
+ \xi_{\phi H} \mu_H^2 \right) = \frac{c^2 }{a^6 }
\,.
\end{eqnarray}
The effective potential after eliminating $H_1$ is given by
\begin{eqnarray}
&&
V_{\rm eff}( y) = \frac{\xi_{\phi H}}{2} \mu_H^4
\left(  \epsilon y^2 + 2 y + \frac{2 \delta_c}{y}
\right)
\,, 
\\ &&
y \equiv \frac{ \phi_t^2}{\mu_H^2}
\,, \hspace{0.5cm}
\epsilon \equiv - \frac{\xi_{\phi H}^2 - 4 \lambda_{\phi} \lambda_H}{2 \lambda_H \xi_{\phi H} } 
\,, \hspace{0.5cm}
\delta_c \equiv  
\frac{1 }{2\xi_{\phi H} }(\frac{c }{a^3 \mu_H^3 })^2 
\,.
\end{eqnarray}

An interesting case is given when two terms cancels such that
$ \partial_y ( \epsilon y^2  + \frac{2 \delta_c}{y}) \approx 0 $
under the condition $4\lambda_{\phi}\lambda_H > \xi_{\phi H}^2 $.
The tracker solution in this case is
\begin{eqnarray}
&&
\phi_t = 3^{1/4} 2^{7/12} \lambda_H^{1/6} (\frac{4 \lambda_H }
{- \xi_{\phi H}^2 + 4\lambda_{\phi}\lambda_H  })^{1/12}
\frac{ (H_0 M_{\rm P})^{1/2}}{a(t)}
\,,
\\&&
V_{\rm eff} = 6 (- \frac{\xi_{\phi H}^2 }{4\lambda_H }+\lambda_{\phi} )^{2/3} 
 \frac{(H_0 M_{\rm P})^2}{a(t)^4}
\,, \hspace{0.5cm}
a(t) = \frac{t+ t_i}{t_0 + t_i}
\,.
\end{eqnarray}
Mass squared matrix for dark matter fields is given by
\begin{eqnarray}
&&
(\delta \phi_1, \delta \phi_t)\overline{ {\cal M}^2} 
\left(
\begin{array}{c}
\delta \phi_1  \\
\delta \phi_t   
\end{array}
\right)
\,, \hspace{0.5cm}
\overline{ {\cal M}^2} = \left(
\begin{array}{cc}
4 \lambda_{\phi} \phi_t^2
& 0 \\
 0
&  (12 \lambda_{\phi} + \frac{3 }{\lambda_H }(1- \xi_{\phi H} )\, )\phi_t^2 
\end{array}
\right)
\,.
\label {mass sq matrix 2}
\end{eqnarray}
This model describes a dark matter mass much closer to standard particle physics,
of order $ \lambda_{\phi}^{2/3} \times $ a few meV.

The full set of non-linear coupled equation using dimensionless
fields $Z=\phi_t/\mu_H$ and $U=\phi_1/\mu_H$ is given by
\begin{eqnarray}
&&
\frac{d^2 Z}{d \tau^2} + \frac{3}{a} \frac{da }{ d\tau } \frac{d Z}{d \tau} +  
\partial_Z v_{\rm eff}(Z, U) = 0
\,,
\label {o3 eq1}
\\ &&
\frac{d^2 Z}{d \tau^2} + \frac{3}{a} \frac{da }{d\tau } \frac{d Z}{d \tau} +  
\partial_U v_{\rm eff}(Z, U) = 0
\,,
\label {o3 eq2}
\\ &&
( \frac{da }{a d\tau })^2 = \frac{ 1}{ 6} b^2
\left( \frac{1}{2} (\frac{dZ }{d\tau })^2 + \frac{1}{2} (\frac{dU }{d\tau })^2
+ v_{\rm eff}(Z, U)
\right)
\,,
\label {o3 eq3}
\\ &&
\hspace*{-1cm}
v_{\rm eff}(Z, U) = \frac{ \xi_{\phi H}}{2} ( \epsilon Z^4 + 2 Z^2 + \frac{2 \delta_c}{Z^2}
+ 2 \epsilon Z^2 U^2 + \delta_{\lambda} U^4 + 2 U^2 )
\,, \hspace{0.5cm}
b = \frac{\mu_H}{M_{\rm P}}
\,, \hspace{0.5cm}
\delta_{\lambda} = \frac{2 \lambda_{\phi}}{\xi_{\phi H}}
\,.
\label {non-linear coupled-eq}
\end{eqnarray}

\begin{figure*}[htbp]
 \begin{center}
 \epsfxsize=0.6\textwidth
 \centerline{\epsfbox{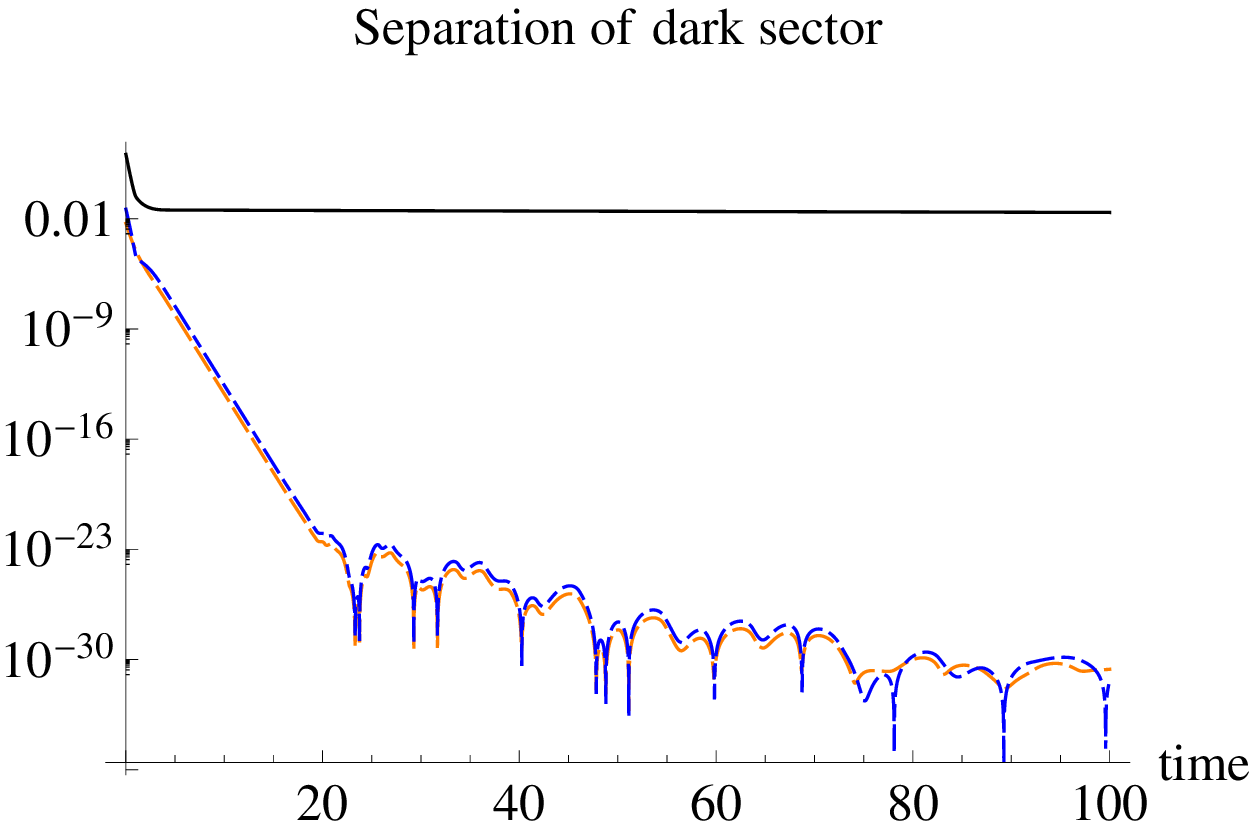}} \hspace*{\fill}\vspace*{1cm}
   \caption{
Example of time evolution by solving the full set of non-linear coupled equation,
eq.(\ref{o3 eq1}) $\sim $ eq.(\ref{o3 eq3}) 
based on the effective potential $v_{{\rm eff}}(Z, U)  $ of eq.(\ref{non-linear coupled-eq})
for parameter choice of $b=1/10^3 \, \epsilon = 2,\, \delta_c = 1/10^2,\, \delta_{\lambda} =1/2,
\, \xi_{\phi H} = 2$ and initial condition, $Z=U= 10,\, a =1/10^2$:
total energy density times the volume factor $a^3(t_f)$ at the end time $t_f$ in solid black,
$U$ energy density times $a^3(t)$  in dashed orange, and
$Z, U$ interaction energy density times $a^3(t)$ in dash-dotted blue.
and volume factor $a^3(t) $ in dotted black.
The volume factor increases by $\sim 4$ in this time range.
}
   \label {dark sector separation}
 \end{center} 
\end{figure*}

We illustrate in Fig(\ref{dark sector separation})
 an example of solution obtained numerically integrating
the full set of non-linear coupled equations, 
eq.(\ref{o3 eq1}) $\sim $ eq.(\ref{o3 eq3}).
Only $\phi_t$ dark energy remains at the end point of computation,
rapidly dumping $\phi_1$ and mutual interaction energy densities.

\subsection
{\bf Bifurcation at electroweak critical temperature}

One has to incorporate effects of high temperature in the early universe.
We are particularly interested in behaviors of order parameters, $H_1, \phi_t,$ around
the electroweak phase transition at temperature $T=T_c$ (electroweak critical temperature).
The effective potential is now replaced by the Gibbs free energy, and
one has a finite temperature contribution to field squared term of the form,
$\eta\, \alpha\, T^2 H_1^2,\, \eta>0 $, with $\eta$ a dimensionless constant of order unity.
We keep using the terminology of effective potential instead of the Gibbs free energy, 
to write it as
\begin{eqnarray}
&&
\hspace*{-1cm}
V_{\rm eff}(H_1, \phi_1, \phi_t; T) =  
\lambda_{\phi} (\phi_1^2 + \phi_t^2 - \frac{m_{\phi}^2}{4 \lambda_{\phi}})^2
+ \lambda_H (H_1^2 - \mu_H^2 )^2 + \eta \alpha\, T^2 H_1^2
 + \xi_{\phi H}\,(\phi_1^2 + \phi_t^2)  H_1^2  
+ \frac{c^2}{ 2 a^6 \phi_t^2}
\,.
\label {new higgs potential 2}
\end{eqnarray}
There is no temperature dependent correction related to scalar field $\phi$
due to their extremely feeble interaction of gravitational strength with ambient particles.

The equation that determines stationary points is modified by temperature dependent term;
below $T < T_c $ stationary points are determined by
\begin{eqnarray}
&&
H_1^2 = \mu_H^2 - \frac{\eta \alpha}{2 \lambda_H} T^2 
- \frac{\xi_{\phi H}}{2 \lambda_H} \phi_t^2
= \frac{\eta \alpha}{2 \lambda_H} (T_c^2 - T^2) - \frac{\xi_{\phi H}}{2 \lambda_H} \phi_t^2
\,, \hspace{0.5cm}
T_c = \sqrt{\frac{ 2\lambda_H}{\eta \alpha}}\, \mu_H
\,,
\\ &&
2 \phi_t^4 \left( 2\lambda_{\phi} (1 - \frac{\xi_{\phi H}^2}{4\lambda_{\phi} \lambda_H})\phi_t^2 + 
\frac{\eta \alpha\, \xi_{\phi H}}{2 \lambda_H} (T_c^2 - T^2) 
\right) = \frac{c^2}{a^6}
\,,
\end{eqnarray}
with $\phi_1 = 0$.
The obvious constraint $H_1^2 \geq 0$ gives a condition;
\begin{eqnarray}
&&
\phi_t^2 \leq \frac{\eta \alpha}{ \xi_{\phi H}} ( T_c^2 - T^2)
\,.
\end{eqnarray}
As $T \rightarrow T_c^-$, $\phi_t^2 \rightarrow 0^+$.
Thus, we establish that both of $\phi_1, \phi_t$ vanish above the critical temperature,
and the electroweak critical temperature is the point of bifurcation of
dark scalar fields.
From a consistency of the above equation we conclude that
$c$ also approaches to zero at the critical temperature,
\begin{eqnarray}
&&
c \rightarrow {\rm constant}\, (T_c- T)^{3/2}
\,.
\end{eqnarray}


\subsection
{\bf Doublet case is not viable}

We have also considered the case of dark scalars being SU(2) $\times $ U(1) doublet.
Both complex SU(2) doublets may be described by four real fields, $H_i\,, \phi_i\,, i = 1 \sim 4$,
\begin{eqnarray}
&&
H = \left(
\begin{array}{c}
H_1 + i H_2  \\
H_3+ i H_4   
\end{array}
\right)
\,, \hspace{0.5cm}
\phi = \left(
\begin{array}{c}
\phi_1 + i \phi_2  \\
\phi_3 + i \phi_4   
\end{array}
\right)
\,, \hspace{0.5cm}
\phi^{\dagger} H = \sum_{i=1}^4 H_i \phi_i
\,.
\end{eqnarray}
We take an extended SU(2) $\times$ U(1) symmetric  potential replacing the usual Higgs potential by
\begin{eqnarray}
&&
V_{\phi H} =  \lambda_{\phi} (\phi^{\dagger}\phi - \frac{m_{\phi}^2}{2 \lambda_{\phi}})^2
+ \lambda_H (H^{\dagger} H - \mu_H^2 )^2 
+ \lambda_{\phi H} (\phi^{\dagger} H - \mu_{\phi H}^2 )^2
+ \xi_{\phi H} \,\phi^{\dagger}\phi H^{\dagger} H
\,,
\label {new higgs potential}
\end{eqnarray}
with $ m_{\phi}^2\,,\lambda_{\phi}\,, \lambda_H\,, \lambda_{\phi H}\,, \xi_{\phi H}$ 
all taken positive.

We studied this model in great detail, but the bottom line is
existence of stable charged scalars $\phi^{\pm}$ of masses, $O$(a few meV).
The charged scalars have electromagnetic interaction, and their pair production
via analogue of Bethe-Heitler pair production for $e^{\pm}$ 
is our concern. Cross section is lots larger by the factor $(m_e/m(\phi^{\pm})\,)^2$
than Bethe-Heitler process, and it drastically 
modifies the behavior of electromagnetic showers above
the shielding energy $\approx $ keV.
We believe that this is sufficient to reject the doublet case.

\section
{\bf Coupling of neutral dark scalar to matter and laboratory test}

An attractive feature of presented models is the uniqueness
of coupling to standard theory particles:  the new degree of scalar field
couples to standard particles by the conformal factor $1/f^2$
 in the Einstein frame lagrangian, 
and $- (\partial_{\phi} f/ 2f )T^m $ in the field equation.
This coupling is  related to cosmological evolution of scalar fields
via varying $f(\phi)$, giving prediction of testable consequences.

First of all, the tracker solution at recent epochs 
predicts variation of basic coupling constants.
For the fine structure constant, its fractional variation is found to be
of order $H_0/M_{\rm P} $ divided by the age of universe
in both viable models of Section 3 and Section 4.
For instance, in genetic model of Section 3 it is given by
\begin{eqnarray}
&&
| \frac{\dot{\alpha}}{\alpha} | = 4 |\frac{d}{dt} \ln \frac{1}{f} |
= \frac{2^{7/4} }{5} f_2 (\frac{ \lambda}{25 })^{-1/2 } 
\frac{H_0 }{M_{\rm P} } \frac{1 }{H_0 (t+t_i)^2 }
\,, \hspace{0.5cm}
2^{7/4} \frac{H_0}{M_{\rm P} t_0} 
\sim 2 \times 10^{-70}\, {\rm year}^{-1}
\,,
\label {alpha variation}
\end{eqnarray}
which is utterly impossible to detect.

For discussion of model tests in laboratories on earth
we recapitulate  in the following table results on dark energy density,
masses of linearized quanta, and
their coupling to standard model particles as defined by the  coefficient of $-T^m$, which was
 obtained in the preceding section.
The required minimal number of parameters, hence the most predictive
results are assumed in the table, hence these numbers are only a guide ignoring
more complexities.

\vspace{0.5cm}
\hspace*{-1cm}
\begin{tabular} {cccc}
Models & Dark energy density & Dark quantum mass 
& Coupling to standard particles  \\ \hline \hline
O(2) & $ (1.7 {\rm meV})^4 $ &  $1.3 \lambda^{1/4} \, {\rm meV}$
& $\sqrt{\frac{f_2 }{ 5}} \frac{1}{M_{\rm P}}   $   \\ \hline
SU(2) $\times $ U(1) singlet bifurcation & $ (1.7 {\rm meV})^4 $  
& $O(0.95 {\rm meV\; and\;} \, 1.6 {\rm meV})$
&$ \sqrt{\frac{f_2 }{ 5}} \frac{1}{M_{\rm P}} $  \\ \hline
\end{tabular}
\vspace{0.5cm}

We have studied possibilities for dark quantum search in table-top
 atomic experiments using coherence, but have found no good method.
The best experimental method seems to be the classical approach of the firth-force search
such as refined torsion balance experiments.
If one could focus on the force range of scalar mediated interaction, $O$(0.1) mm,
corresponding to a few meV mass,
in these experiments, the method may become ideal.

Neither result of optical rotation \cite{ad search optical} 
nor of light shining through wall method \cite{ad search shining wall}
is effective to constrain our model, since these experiments 
are sensitive to the effective two-photon coupling of order $> 10^{-6} \sim 10^{-7} \,$GeV$^{-1}$, 
while our corresponding coupling is of order 1/the Planck energy.

\section
{\bf Summary}

The multi-scalar field $\phi=(\phi_i)\,, i=1, \cdots N$ is 
a key for realization of slow-roll inflation and accelerating universe at late times.
Conformal coupling $f(\phi)$ to the scalar curvature $R$ in lagrangian provides to angular
Nambu-Goldstone modes a centrifugal repulsive potential, a positive constant times 
$1/(- g\, \phi_t^2)$ where $\phi_t^2= \sum_{i=1}^n\,, n \leq N$ is the modulus of scalar field.
This repulsive force shifts  potential minimum location $\phi_t$ with cosmological evolution,
 otherwise fixed by an original fundamental lagrangian.
Resulting tracker solution recognized as spatially homogeneous component
 follows a time varying potential minimum  at late times, and gives rise  to 
dark energy density of order $(H_0 M_{\rm P})^2 = $(a few meV)$^{4}$,
with the equation of state factor -1.
On the other hand, the spatially inhomogeneous component gives a dark matter
candidate, its number density being
inversely proportional to the volume factor, third power of cosmic scale factor.
The scalar boson mass is predicted to be an order unity coupling constant times 
$(H_0 M_{\rm P})^{1/2} \approx \,$meV.

We further proposed a mechanism of  bifurcated symmetry breaking 
triggered by the electroweak SU(2) $\times $ U(1) gauge symmetry breaking.
It was found that viable models preclude SU(2) $\times $ U(1) quantum numbers
for dark scalar.
Nambu-Goldstone modes emerge as a result of bifurcation
of symmetry breaking.

All models discussed in the present work pass tests of general relativity at small
cosmological distances, yet it deviates from general relativity
at cosmological scales.
Experimental method to search for dark matter quantum 
in laboratories on earth has also been discussed.
Considering $10^{-2}$ cm range corresponding to 1 meV,
search for the fifth-force type interaction of gravitational strength may be promising.

\vspace{1cm}
 {\bf Acknowledgements}

This research was partially
 supported by Grant-in-Aid   21K03575   from the Japanese
 Ministry of Education, Culture, Sports, Science, and Technology.

\end{document}